\documentclass[nofootinbib,amsmath,amssymb,prb,notitlepage,twocolumn]{revtex4}

\usepackage{graphicx}			% Include figure files
\usepackage{dcolumn}			% Align table columns on decimalpoint
\usepackage{bm}					% bold math
\usepackage{hyperref}			% add hypertext capabilities
\usepackage{amsmath}
\usepackage{amsfonts}
\usepackage{amssymb}
\usepackage{pstcol,times,color,graphicx,graphics,pstricks,pst-node,pst-coil,pst-grad}
\begin{document}
\title{Quantum two-photon emission in a photonic cavity} 
	
% % % % % % % % % % % % % % % % % % % % % %

\author{Y. Muniz}
\email{yurimuniz@pos.if.ufrj.br}
\affiliation{Instituto de F\'{\i}sica, Universidade Federal do Rio de Janeiro, Caixa Postal 68528, Rio de Janeiro 21941-972, RJ, Brazil}

\author{D. Szilard}
\email{daniela@if.ufrj.br}
\affiliation{Centro Brasileiro de Pesquisas F\'isicas, Rua Dr. Xavier Sigaud, 150, 22.290-180, Rio de Janeiro-RJ, Brazil}
	
\author{W. J. M. Kort-Kamp}
\email{kortkamp@lanl.gov}
\affiliation{Theoretical Division, Los Alamos National Laboratory, MS B262, Los Alamos, New Mexico 87545, United States}	
	
\author{F. S. S. da Rosa}
\email{frosa@if.ufrj.br}
\affiliation{Instituto de F\'{\i}sica, Universidade Federal do Rio de Janeiro, Caixa Postal 68528, Rio de Janeiro 21941-972, RJ, Brazil}

\author{C. Farina}
\email{farina@if.ufrj.br}
\affiliation{Instituto de F\'{\i}sica, Universidade Federal do Rio de Janeiro, Caixa Postal 68528, Rio de Janeiro 21941-972, RJ, Brazil}

% % % % % % % % % % % % % % % % % % % % % % 	

%\author[1]{Y. Muniz}
%\author[2]{D. Szilard}
%\author[3]{W.J.M. Kort-Kamp}
%\author[1]{F.S.S. Rosa}
%\author[1]{C. Farina}

%\affil[1]{XXX}
%\affil[2]{ZZZ}
%\affil[3]{AAA}
%\date{\today}

\begin{abstract}

We  derive a new expression for the two-photon spontaneous emission (TPSE) rate of an excited quantum emitter in the presence of arbitrary bodies in its vicinities. After investigating the influence of a perfectly conducting plate on the TPSE spectral distribution (Purcell effect), we demonstrate the equivalence of our expression with the more usual formula written in terms of the corresponding dyadic Green's function. We establish a convenient relation between the TPSE spectral distribution and the corresponding Purcell factors of the system. Next, we consider an emitter close to a dielectric medium and show that, in the near field regime, the TPSE spectral distribution is substantially enhanced and changes abruptly at the resonance frequencies. Finally, motivated by the suppression that may occur in the one-photon spontaneous emission of an excited atom between two parallel conducting plates, we discuss the TPSE for this same situation and show that complete suppression can never occur for $s \rightarrow s$ transitions.

\end{abstract}

\maketitle

\section{Introduction}

Just a few years after Dirac's seminal paper on Quantum Electrodynamics (QED)\cite{dirac1927}, M. G{\"o}ppert-Mayer investigated elementary processes involving two quantum transitions\cite{goppert1931}. Since then, these processes have attracted the attention of many researchers and have shown to be relevant in many branches of physics.  An important example is the role played by the two-photon absorption process in the study  of the 1S-2S transition in hydrogen.  The corresponding transition frequency has been measured recently with an accuracy of a few parts in $10^{15}$, a remarkable achievement \cite{parthey2011}. This kind of investigation is relevant, for instance, in setting limits on possible time variations of fundamental constants \cite{fischer2004} or in the investigation of possible violations of Lorentz invariance \cite{altschul2010}. Two-photon absorption processes are equally important in the study of the 1S-2S transition in antihydrogen \cite{ahmadi2018}, since its comparison with the same transition in hydrogen provides valuable information to test charge, parity, and time reversal (CPT) symmetry  \cite{gabrielse1987,andresen2011}.

In this work, instead of two-photon absorption processes,  we shall be concerned with the calculation of two-photon spontaneous emission (TPSE) rates. Though the theoretical developments of such processes had already been made in the pioneering paper by G{\"o}ppert-Mayer\cite{goppert1931}, it took approximately one decade until Breit and Teller made explicit numerical estimates  for the TPSE rates in hydrogen and helium \cite{breit1940}. Another decade was necessary until TPSE was used to give the correct explanation of the continuous emission of planetary nebulae, presented  by Spitzer and Greenstein in 1951 \cite{spitzer1951}. Subsequently, many theoretical papers were published in the field with better numerical calculations of the TPSE rate for hydrogenic atoms\cite{shapiro1959}. However, only in 1965 a first direct measurement of this phenomenon was performed by Lipeles \textit{et al} using ionized Helium\cite{lipeles1965}, and an experiment with hydrogen atoms came up only in 1975\cite{kruger1975}. Since then, TPSE has been investigated in many other systems, such as many-electron atoms\cite{bannett1982}, semiconductors\cite{hayat2007,hayat2008,driel2008} and quantum dots\cite{ota2011}. Recently, due to the new technological developments and the growing progress in near-field optics, plasmonics, and materials science in general, TPSE has attracted new interests. These are mainly connected to the possibility of controlling TPSE generation and its properties with external agents\cite{hayat2010,kivshar2012},  in what could be called a two-photon Purcell effect\cite{purcell1946}.

Since TPSE plays a key role in many important topics, such as quantum cryptography and computing, controlling it is of extreme interest both theoretically and experimentally. In recent work, it has been shown that TPSE processes may occur at very short time scales in comparison to those of conventionally fast transitions \cite{rivera2016} and can even dominate the one-photon spontaneous emission (SE)\cite{rivera2017}. For instance, due to the high confinement offered by phonon polariton modes of a polar dielectric over a sufficiently narrow frequency band, an excited emitter prefers to decay via the simultaneous emission of two quanta, despite the possibility of allowed single-photon decay pathways.

In this work we have two main purposes: a general theoretical discussion of the two-photon Purcell effect and the study of this phenomenon in some simple but interesting systems. Initially, we establish a new formula for computing the Purcell effect in TPSE processes. In this framework, the TPSE rate is written in terms of the electromagnetic field modes satisfying the boundary conditions imposed by the environment of the emitter. There are many motivations for doing that. First of all, such a formula for TPSE was lacking in the literature, despite the fact that an approach based on the field modes can be very useful in situations where they are explicitly known, as for instance, in multilayered media\cite{aleksander1995,reyes2009,contreras2009,eberlein2010}, photonic crystals\cite{joannopoulos2008,li2000,johnson1999}, among others\cite{wu1999,klimov2004,eberlein2009,kort2013}. Moreover, this approach is more suited  for computing the angular distribution of the emitted photons when compared to the usual one based on the Green's function. We illustrate this formula by considering an emitter close to a perfectly conducting plate. We then demonstrate its equivalence with the Green's function method and compare both methods by reobtaining the TPSE rate of the emitter-mirror system. We also establish a connection between the TPSE rate of an emitter near an arbitrary object and the one-photon SE Purcell factors in the same situation. As we will show, this allows one to calculate straightforwardly  the TPSE rate once the corresponding one-photon SE rates are known.  With this approach, we investigate the TPSE of an emitter near a semi-infinite homogeneous dielectric material as well as the case of an atom between two perfectly conducting plates.  For the latter case, we show that, in contrast to the one-photon SE, complete suppression of the TPSE can never occur in this situation. 
 
 This paper is organised as follows. In Section II we establish the formula for the TPSE rate based on the electromagnetic field modes and illustrate it in a specific example. In Section III we demonstrate its equivalence with the Green's function method and, with the purpose of comparing both of them, we reobtain the same results calculated in the previous section. In Section IV we show how the TPSE rate can be written in terms of the one-photon SE Purcell factors. In Section V we consider a quantum emitter inside an open cavity formed by two parallel mirrors and show that suppression of TPSE can never occur for $s\rightarrow s$ transitions. Section VI is left for final remarks and conclusions. For convenience, an appendix was included.

\section{TPSE rate: field modes approach}
Here we present a simple way to compute the TPSE rate of an atom close to a surface of arbitrary shape. The Hamiltonian of the system  is given by  $H = H_{A} + H_F + H_{int}$, where $H_{A}$ is the emitter's hamiltonian, $H_F$ is the field hamiltonian, and $H_{int}$ describes the emitter-field interaction, to be treated by perturbative methods. The second-order transition rate between an initial state $|i\rangle$ and a final state $|f\rangle$ is given by Fermi's golden rule \cite{sakurai2014,thiru1984},
\begin{equation}
\Gamma_{i \rightarrow f} = \frac{2\pi}{\hbar}|M_{fi}|^2\delta(E_f - E_i)\, ,
\end{equation}
where
\begin{equation}
M_{fi} = \sum_I \frac{\langle f|H_{int}|I\rangle\langle I|H_{int}|i\rangle}{E_i - E_I}\, .
\end{equation}
The dominant  transition wavelengths are assumed to be much larger than the emitter dimensions, so that the electric dipole approximation is valid and we can write\cite{milonni2013}
\begin{eqnarray}
H_{int} \!\!&=& \!\!-{\bf d}\cdot{\bf E}({\bf r}) \cr
&=& \!\!-i \!\sum_\alpha \!\!\sqrt{\frac{\hbar\omega_\alpha}{2\epsilon_0}}\!\left[a_\alpha {\bf d}\cdot{\bf A}_\alpha({\bf r}) \! -\!  a^\dagger_\alpha {\bf d}\cdot{\bf A}^\ast_\alpha({\bf r})\right]\! ,
\end{eqnarray}
where ${\bf d}$ is the dipole moment operator, ${\bf r}$ is the emitter's position and $\{{\bf A}_\alpha\}$ is a complete set of solutions of the Helmholtz equation subjected to the boundary conditions imposed by the surface. Initially, the emitter is in an excited state and the field in the vacuum state, so that the emitter-field initial state is written as $|i\rangle = |e;0\rangle$. Since we are interested in TPSE processes, the final state must be constituted by the emitter in a lower energy state and the field in a two-photon state, namely,  $|g;1_\alpha , 1_{\alpha'}\rangle$. The intermediate states $|I\rangle$ that connect the initial and final states are $|m;1_\alpha\rangle$ or $|m;1_{\alpha'}\rangle$, where $m$ indexes the emitter states. Summing over all possible final states and defining 
\begin{equation}
\mathbb{D}(\omega_{\alpha},\omega_{\alpha'}) := \sum_m\left[\frac{{\bf d}_{em}{\bf d}_{mg}}{\omega_{em} - \omega_{\alpha}} + \frac{{\bf d}_{mg}{\bf d}_{em}}{\omega_{em} - \omega_{\alpha'}}\right], \label{diadico}
\end{equation}
where ${\bf d}_{ab} := \langle a|{\bf d}| b\rangle$ and $\omega_{ab} := \frac{E_a - E_b}{\hbar}$, we obtain
\begin{widetext}
\begin{equation}
\Gamma({\bf r}) = \frac{\pi}{4\epsilon_0^2\hbar^2} 
\sum_{\alpha,\alpha'}\omega_{\alpha}\omega_{\alpha'}|{\bf A}_{\alpha}({\bf r})\cdot \mathbb{D}(\omega_{\alpha},\omega_{\alpha'})
 \cdot {\bf A}_{\alpha'}({\bf r})|^2\delta(\omega_{\alpha} + \omega_{\alpha'} -\omega_{eg}). 
\label{GammaModos}
\end{equation}
\end{widetext}
This formula can be viewed as a second-order equivalent of
\begin{equation}
\Gamma^{(1)}({\bf r}) = \frac{\pi}{\epsilon_0\hbar}\sum_{\alpha}\omega_{\alpha}|{\bf d}_{eg}\cdot{\bf A}_{\alpha}({\bf r})|^2\delta(\omega_{\alpha} - \omega_{eg})\, , \label{Gamma1Modos}
\end{equation}
which can be found in the literature\cite{milonni2013,novotny2012,kort2013} and gives the one-photon SE rate of an atom near an arbitrary surface.

As an immediate application of Eq. \eqref{GammaModos},  we shall first reobtain  the  TPSE rate in free space\cite{goppert1931,thiru1984}. In this case, the field modes $ {\bf A}_\alpha({\bf r})$ can be chosen as $ {\bf A}_{{\bf k},\lambda}({\bf r}) = \frac{e^{i{\bf k}\cdot{\bf r}}}{\sqrt{V}}\boldsymbol{\epsilon}_{{\bf k}\lambda}$, where $V$ is a box quantization volume and $\{\boldsymbol{\epsilon}_{{\bf k}\lambda}; \lambda = 1,2\}$ are the unit polarization vectors. Taking the limit to the continuum, we make the replacement $\sum_\alpha \rightarrow \frac{V}{(2\pi)^3}\sum_\lambda\int d^3{\bf k}$, and find
\begin{equation}
\Gamma_0 = \int_0^{\omega_{eg}}d\omega\,\gamma_0(\omega),
\end{equation}
where $\gamma_0$ is the free space spectral distribution of the emitted photons, given by
\begin{equation}
\gamma_0(\omega) = \frac{\mu_0^2}{36\pi^3\hbar^2c^2}\omega^3(\omega_{eg} - \omega)^3\vert\mathbb{D}(\omega,\omega_{eg} - \omega)\vert^2\, ,
\end{equation}
with $\vert\mathbb{D}(\omega,\omega_{eg} - \omega)\vert^2 := \mathbb{D}_{ij}(\omega,\omega_{eg} - \omega)\mathbb{D}^\ast_{ij}(\omega,\omega_{eg} - \omega)$. Essentialy, $\gamma_0(\omega)d\omega$ gives the number of emitted photons per unit time in the interval $[\omega, \omega + d\omega]$. A few  comments are in order here. First, due to the homogeneity of space, note that $\Gamma_0$ is independent of the position of the emitter. Second,  in contrast to the one-photon SE, which is a narrow band phenomenon, TPSE is a broadband phenomenon. Another important feature of the above spectral distribution is its symmetry with respect to $\omega_{eg}/2$, namely, $\gamma_0(\omega) = \gamma_0(\omega_{eg} - \omega)$, which is a direct consequence of energy conservation since the sum of the frequencies of the two emitted photons must equal the emitter's transition frequency (recoil is being neglected in the present calculations).
\subsection{An emitter near a perfectly conducting plate}
As another application of Eq. \eqref{GammaModos}, let us compute the Purcell effect in the TPSE rate by considering an emitter separated by a distance $z$ from a perfectly conducting plate placed at ${z = 0}$. The electromagnetic field modes satisfying the boundary conditions ${{\bf E}\times{\bf \hat{z}}\big\vert_{z=0} = 0}$ and ${{\bf B}\cdot{\bf \hat{z}}\big\vert_{z=0} = 0}$ are given by\cite{milonni2013,milonni1973}
\begin{eqnarray}
{\bf A}_{{\bf k},1}({\bf r}) &=& \sqrt{\frac{2}{V}}\sin(k_zz)e^{i{\bf k}_\parallel\cdot{\bf r}}({\bf \hat{k}}_\parallel\times{\bf \hat{z}}), \label{Modo1}
\\
{\bf A}_{{\bf k},2}({\bf r}) &=& \sqrt{\frac{2}{V}}\frac{1}{k}[k_\parallel\cos(k_zz){\bf \hat{z}} - ik_z\sin(k_zz){\bf \hat{k}}_\parallel]e^{i{\bf k}_\parallel\cdot{\bf r}}.\cr
&{\,}& \label{Modo2}
\end{eqnarray}
Substituting Eqs. (\ref{Modo1}) and (\ref{Modo2}) into Eq. (\ref{GammaModos}) and performing the integrals in the azimuthal angles $\phi$ and $\phi^\prime$ we obtain 
\begin{equation}
\Gamma(z) = \int_0^{\omega_{eg}}d\omega\int_0^\pi d\theta d\theta^\prime \,S(\omega,\theta,\theta^\prime;z)\, ,
\end{equation}
where $S(\omega,\theta,\theta^\prime;z)$ is the angular distribution of emitted photons with respect to the $z$ axis and is given by
%
%\begin{equation}
%S = S_\parallel + S_\perp + S_c\, ,
%\end{equation}
%
%where
%
%
\begin{widetext}
\begin{eqnarray}
S &=& S_\parallel + S_\perp + S_c,
\\
\mbox{where} &{\;}& \cr
S_\parallel &=& \frac{\mu_0^2}{64\hbar^2\pi^3}\omega^3(\omega_{eg} - \omega)^3\sum_{i,j =1,2}|\mathbb{D}_{ij}(\omega,\omega_{eg} - \omega)|^2\left[ \sin^2(kz\cos\theta)\sin\theta(1 + \cos^2\theta)\right]\left[ \sin^2(kz\cos\theta^\prime)\sin\theta^\prime(1 + \cos^2\theta^\prime)\right], \cr
&{\,}&
\\
S_\perp &=& \frac{\mu_0^2}{16\hbar^2\pi^3}\omega^3(\omega_{eg} - \omega)^3|\mathbb{D}_{33}(\omega,\omega_{eg} - \omega)|^2\cos^2(kz\cos\theta)\sin^3\theta\cos^2(kz\cos\theta^\prime)\sin^3\theta^\prime ,
\\
S_c &=& \frac{\mu_0^2}{32\hbar^2\pi^3}\omega^3(\omega_{eg} - \omega)^3\sum_{i = 1,2}\bigg\{|\mathbb{D}_{i3}(\omega,\omega_{eg} - \omega)|^2\left[ \sin^2(kz\cos\theta)\sin\theta(1 + \cos^2\theta)\right]\cos^2(kz\cos\theta^\prime)\sin^3\theta^\prime +\cr
&+& |\mathbb{D}_{3i}(\omega,\omega_{eg} - \omega)|^2\cos^2(kz\cos\theta)\sin^3\theta\left[\sin^2(kz\cos\theta^\prime)\sin\theta^\prime(1 + \cos^2\theta^\prime)\right]\bigg\}.
\end{eqnarray}
\end{widetext}
For an observer in the far field, the angular distribution gives the number of emitted photon-pairs with frequencies between $\omega$ and $\omega + d\omega$ and within the angular intervals $[\theta,\theta + d\theta]$ and $[\theta^\prime, \theta^\prime + d\theta^\prime]$. Integrating the angular distribution in $\theta$ and $\theta^\prime$ it is straightforward to show that the TPSE rate can be written in the form
\begin{equation}
\Gamma(z) = \int_0^{\omega_{eg}}d\omega\,\gamma(\omega;z)\, ,
\end{equation}
where the spectral distribution is now given by
%
%\begin{equation}
\begin{align}
\gamma(&\omega;z) = \gamma_0(\omega) \times\cr
&\!\!\times\!\!\sum_{i,j}\! \frac{|\mathbb{D}_{ij}(\omega,\omega_{eg} - \omega)|^2}{|\mathbb{D}(\omega,\omega_{eg} - \omega)|^2}
P_i(\omega;z)P_j(\omega_{eg} - \omega;z) ,
\label{gamma-1placa}
%\end{equation}
\end{align}
with
%
%\begin{eqnarray}
\begin{align}
P_1(\omega&; z) = P_2(\omega;z) := \frac{3}{2}\times\cr
&\times\left[\frac{2}{3} - \frac{\sin(2kz)}{(2kz)} - \frac{\cos(2kz)}{(2kz)^2} + \frac{\sin(2kz)}{(2kz)^3}\right]\, ,
 \label{Pparallel-1placa}
\\
P_3(\omega&; z) := 3\left[\frac{1}{3} - \frac{\cos(2kz)}{(2kz)^2} + \frac{\sin(2kz)}{(2kz)^3}\right]\, , 
\label{Pperp-1placa}
%\end{eqnarray}
\end{align}
being the Purcell factors associated with the one-photon SE of an emitter with a transition dipole moment oriented parallel or perpendicular to the plate, respectively\cite{morawitz1969,milonni2013}, and $k = \omega/c$. Since the presence of the plate breaks the translational symmetry along the $z$ direction, the spectral distribution function depends not only on the frequency but also on the distance from the emitter to the surface.

In a wide variety of two-photon transitions, the initial and final atomic states have spherical symmetry. This is due to the fact that one-photon electric dipole transitions between $s$ states are forbidden by selection rules, and in usual experimental conditions other two-photon transitions are in disadvantage with respect to single photon ones.

%
%\begin{align}
%\gamma(\omega;z) &= \frac{\mu_0^2}{36\pi^3\hbar^2c^2}\bigg\vert\sum_nd_{en}d_{ng}\bigg\vert^2\omega^3(\omega_{eg} - \omega)^3 \times \cr
%&
%\times \bigg\vert\frac{1}{\langle\omega_{en}\rangle - \omega} + \frac{1}{\langle\omega_{en}\rangle - (\omega_{eg} - \omega)}\bigg\vert^2\times\cr\cr
%&
%\times\sum_{i}P_i(\omega;z)P_i(\omega_{eg} - \omega;z) \label{gamma-plot}
%\end{align}
%
%\begin{widetext}
%\begin{equation}
%\gamma(\omega;z) = \frac{\mu_0^2}{36\pi^3\hbar^2c^2}\bigg\vert\sum_nd_{en}d_{ng}\bigg\vert^2\omega^3(\omega_{eg} - \omega)^3 \bigg\vert\frac{1}{\langle\omega_{en}\rangle - \omega} + \frac{1}{\langle%\omega_{en}\rangle - (\omega_{eg} - \omega)}\bigg\vert^2 
%\sum_{i}P_i(\omega;z)P_i(\omega_{eg} - \omega;z) \label{gamma-plot}
%\end{equation}
%\end{widetext}
%

%
\begin{figure}[h!]
\begin{center}
\includegraphics[width=0.8\linewidth]{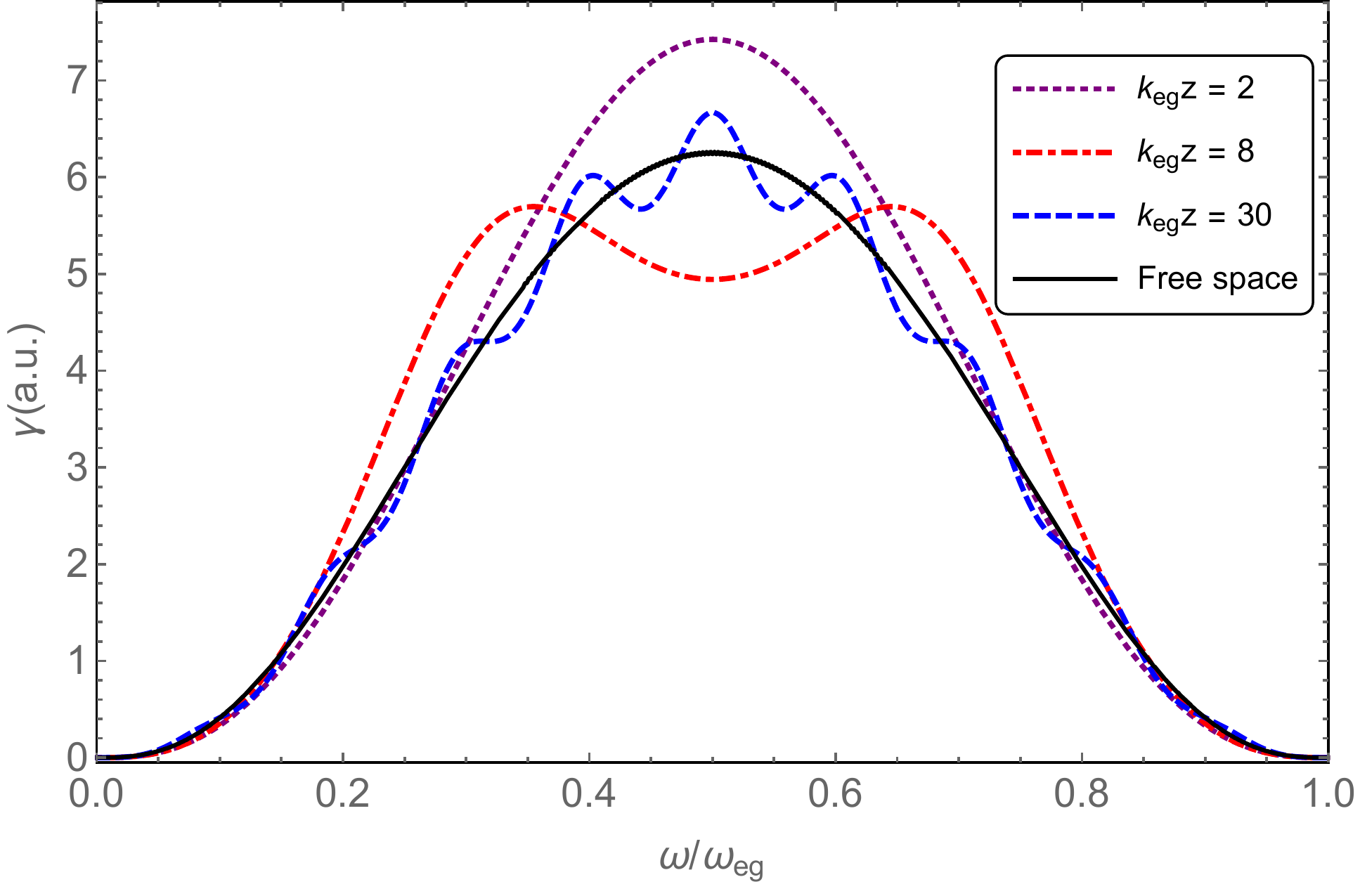}
\caption{Spectral density function $\gamma(\omega;z)$ of an emitter near a perfect mirror in terms of the dimensionless variable $\omega/\omega_{eg}$ 
for three different values of $z$. We also plot the spectral density function in free space (solid line).}
\label{1placafrequencia}
\end{center}
\end{figure}

In Figure \ref{1placafrequencia} we plot $\gamma(\omega;z)$  {\it versus} $\omega$ for different distances between the emitter and the plate. We considered a $s \rightarrow s$ transition and only one intermediate state in Eq. \eqref{diadico}. As expected, note the symmetry of all spectral distributions with respect to $\omega_{eg}/2$. Observe, also, that the spectral distribution may acquire forms quite different from the parabolical shape of the corresponding one in free space and its local maxima may not necessarily occur at $\omega_{eg}/2$. Moreover, as the distance between the emitter and the plate increases, the spectral distribution tends to the free space one.

\begin{figure}[h!]
\begin{center}
\includegraphics[width=0.8\linewidth]{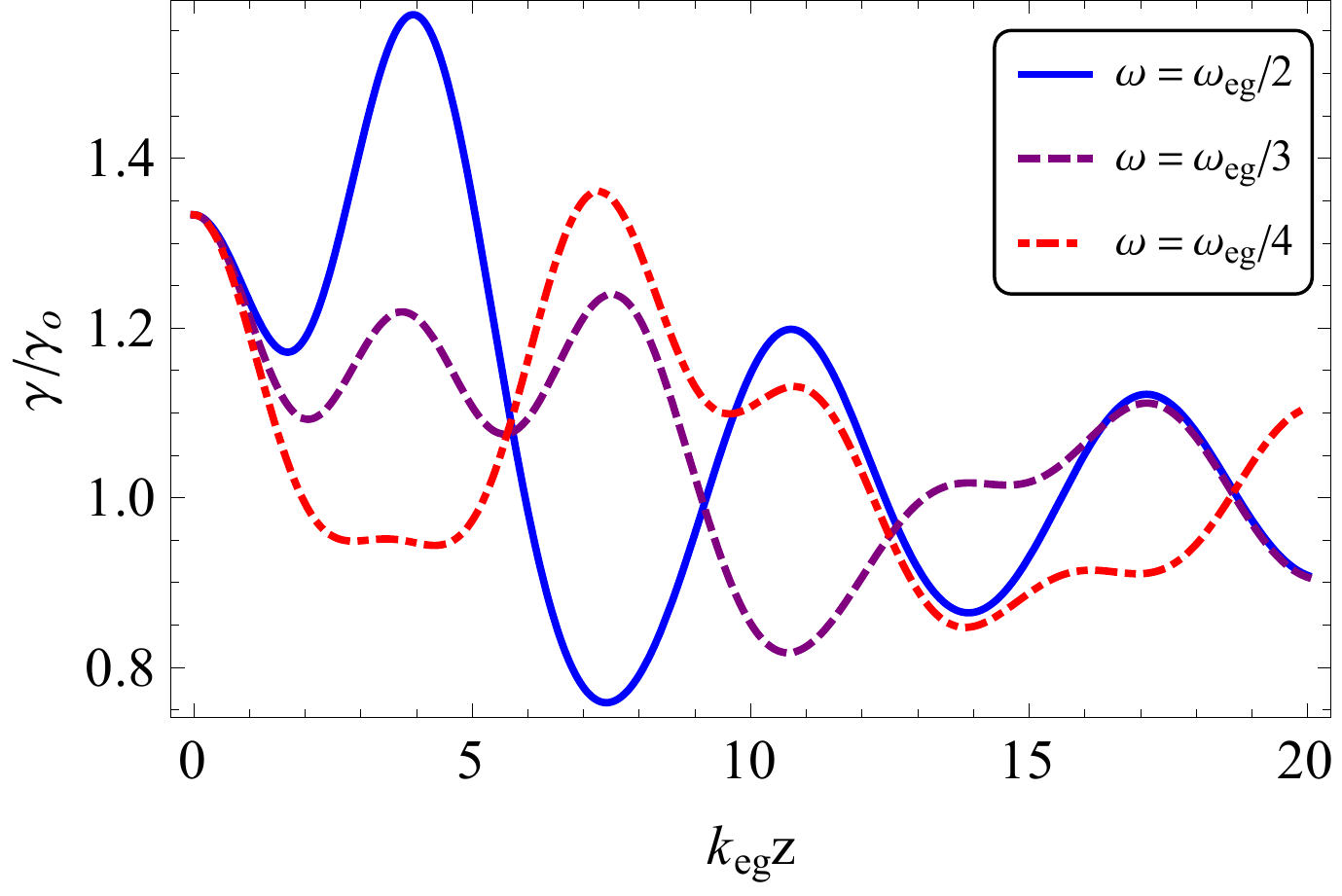}
\caption{Spectral enhancement $\gamma(\omega;z)/\gamma_0(\omega)$ of an emitter near a perfect mirror as a function of the separation distance $z$ for three given frequencies.}
\label{1placadistancia}
\end{center}
\end{figure}

In order to consider all the intermediate states in Eq. \eqref{diadico}, it is convenient to work with the ratio $\gamma(\omega;{\bf r})/\gamma_0(\omega)$, sometimes referred to as spectral enhancement \cite{rivera2017}. For $s \rightarrow s$ transitions, we show that (see appendix)
\begin{equation}
\frac{\gamma(\omega,{\bf r})}{\gamma_0(\omega)} = \frac{1}{3}\sum_{i}P_i(\omega,{\bf r})P_i(\omega_{eg} - \omega,{\bf r}). \label{gamma/gamma_0-transicao-s}
\end{equation}
In Figure \ref{1placadistancia} we plot the spectral enhancement as a function of the distance between the emitter and the mirror for three given frequencies.   As in the one-photon SE rate of an emitter near a perfectly conducting plate, $\gamma(\omega;z)$ also exhibits oscillations with the distance between the emitter and the plate. However, since in the TPSE there is an additional length scale, the oscillations are not as regular as in the one-photon SE, except when $\omega = \omega_{eg}/2$, a particular case in which the two emitted photons have the same wavelength.

We finish this section by emphasizing that the TPSE spectral density function was written in terms of the one-photon Purcell factors of an emitter near a perfectly conducting mirror. Although this has been done in this particular case, this can be generalized, as we show in section {\bf IV}.

%%%%%%%%%%%%%%%%%%%

%
\section{Green's function method}

Another scheme for obtaining SE rates is by using the Green's function formalism. It is widely known that the one-photon SE rate of an emitter near any real material can be written as \cite{novotny2012}
\begin{equation}
\frac{\Gamma^{(1)}({\bf r})}{\Gamma^{(1)}_0} = \frac{6\pi c}{\omega_{eg}}{\bf \hat{n}}_{eg}^\ast\cdot\mathrm{Im}
\mathbb{G}(\omega_{eg};{\bf r},{\bf r})\cdot {\bf \hat{n}}_{eg} , \label{Gamma1Green}
\end{equation}
where $\Gamma^{(1)}_0$ is the one-photon SE rate in free space,  ${{\bf \hat{n}}_{eg}:= {\bf d}_{eg}/|{\bf d}_{eg}|}$ and the Green's dyadic $\mathbb{G}$ satisfies the generalized Helmholtz equation
\begin{equation}
\nabla\times\nabla\times \mathbb{G}(\omega;{\bf r},{\bf r}^\prime)  -\frac{\omega^2}{c^2}\mathbb{G}(\omega;{\bf r},{\bf r}^\prime) =  
\mathbb{I} \delta({\bf r} - {\bf r}^\prime)\, , \label{EquationGreen}
\end{equation}
where $\mathbb{I}$ is the unit tensor, subjected to the appropriate boundary conditions. Here we shall rewrite the TPSE rate given by Eq. \eqref{GammaModos} in terms of the Green's function, thus establishing an equivalence between our result and the literarure\cite{kivshar2012,rivera2016}. With this purpose, first we recall that the dyadic Green's function admits a spectral representation where its imaginary part is given by\cite{novotny2012} 
\begin{equation}
\mathrm{Im}\mathbb{G}(\omega ;{\bf r}, {\bf r}') = \frac{\pi c^2}{2\omega}\sum_{\alpha}{\bf A}_{\alpha}({\bf r}){\bf A}_{\alpha}^\ast({\bf r}')\delta(\omega - \omega_\alpha). \label{GreenRepresentation}
\end{equation}
Noting that from Eq. (\ref{GammaModos}) the TPSE rate can be written as
\begin{eqnarray}\label{Gamma}
\Gamma({\bf r}) &=& \frac{\pi}{4\epsilon_0^2\hbar^2}\sum_{\alpha}\omega_{\alpha}(\omega_{eg} - \omega_{\alpha}){\bf A}_{\alpha}({\bf r})
\cdot \mathbb{D}(\omega_{\alpha},\omega_{eg} - \omega_{\alpha}) \cr
&\cdot & \left[\sum_{\alpha'}{\bf A}_{\alpha'}({\bf r}){\bf A}_{\alpha'}^\ast({\bf r})\delta(\omega_{\alpha} + \omega_{\alpha'} -\omega_{eg})\right]\cr
&\cdot & \mathbb{D}^\dagger(\omega_{\alpha},\omega_{eg} - \omega_{\alpha}) \cdot {\bf A}_{\alpha}^\ast({\bf r})
\end{eqnarray}
and using, for convenience, a simplified notation in which the ${\bf r}$ dependence in the arguments of $\Gamma$, $\mathrm{Im}\mathbb{G}$ and 
${\bf A}_{\alpha}$ is implicit, we obtain  from Eqs. (\ref{GreenRepresentation}) and (\ref{Gamma})
\begin{eqnarray}
\Gamma = \frac{1}{2c^2\epsilon_0^2\hbar^2}\sum_{\alpha}\omega_{\alpha}(\omega_{eg} - \omega_{\alpha})^2 \mathbb{D}_{ij}(\omega_{\alpha},\omega_{eg} - \omega_{\alpha}) \cr\cr
\mathbb{D}_{lk}^\ast(\omega_{\alpha},\omega_{eg} - \omega_{\alpha}) \mathrm{Im}\mathbb{G}_{jk}(\omega_{eg} - \omega_{\alpha}) ({\bf A}_{\alpha})_i({\bf A}^\ast_{\alpha})_l .
\end{eqnarray}
Using that $f(\omega_\alpha) = \int_{-\infty}^{\infty} d\omega f(\omega)\delta(\omega - \omega_\alpha)$ we get
%
%holds for any function $f$, we arrive at
%
\begin{eqnarray}
\Gamma = \frac{\mu_0^2}{\pi\hbar^2}\int_0^{\omega_{eg}}\!\!\!\!\! &{\,}& \!\!\!\!\! d\omega\omega^2(\omega_{eg} - \omega)^2\mathrm{Im}\mathbb{G}_{il}(\omega)\mathrm{Im}\mathbb{G}_{jk}(\omega_{eg} - \omega)\times\cr\cr
&{\times}& \mathbb{D}_{ij}(\omega,\omega_{eg} - \omega)\mathbb{D}_{lk}^\ast(\omega,\omega_{eg} - \omega), \label{GammaGreen}
\end{eqnarray}
where we used again Eq. (\ref{GreenRepresentation}) and constrained the limits of integration since $\mathrm{Im}\mathbb{G}(\omega) = 0$ for $\omega < 0$. 
This equation can be viewed as an equivalent of \eqref{Gamma1Green} for the case of TPSE.

Although Eq. \eqref{GammaGreen} has been derived from \eqref{GammaModos}, which depends on the existence of a complete set of field modes, it is completely general and can be used to calculate the TPSE of an atom near any real material. It is important to notice that Eq. \eqref{GammaGreen} also gives a general formula for the spectral density function $\gamma(\omega;{\bf r})$. 
%
%%%%%%%%%%%%%%%
%
\subsection{Green's function approach to the emitter-mirror system}
%
%%%%%%%%%%%%%%%
%
In order to compare the field modes approach with the Green's function method,  we reobtain in this subsection the TPSE spectral density of an emitter close to a perfect mirror. With this purpose, it is convenient to write the Green function as $\mathbb{G}(\omega;{\bf r},{\bf r}^\prime) = \mathbb{G}^{(0)}(\omega;{\bf r},{\bf r}^\prime) + \mathbb{G}^{(sca)}(\omega;{\bf r},{\bf r}^\prime)$, where $\mathbb{G}^{(0)}$ is the free space Green function and $\mathbb{G}^{(sca)}$ is a homogeneous solution of \eqref{EquationGreen} satisfying the appropriate boundary conditions at the conducting surface.
The calculation of the TPSE rate demands only the knowledge of the imaginary part of the Green's function at coincident points (${\bf r}' = {\bf r}$). In this case, the free space Green function
 is given by $\mathrm{Im}\mathbb{G}^{(0)}(\omega;{\bf r},{\bf r}) = (\omega/6\pi c) \mathbb{I}$ and due to the translational symmetry of the system along any direction parallel to the  ${\cal O}xy$ plane, the scattered Green's function can be written as\cite{novotny2012}
\begin{eqnarray}
\mathbb{G}^{(sca)}(\omega;{\bf r},{\bf r})\!\!\! &=& \frac{i}{8\pi k^2}\int_0^\infty dk_\parallel\frac{k_\parallel}{k_z}e^{2ik_zz}\times \cr\cr
\!\!\!\!\! &\times & \!\!\!\!\!\!\begin{bmatrix} 
k^2r^s - k_z^2r^p & 0 & 0 \\
0 & k^2r^s - k_z^2r^p & 0 \\
0 & 0 & k_\parallel^2r^p
\end{bmatrix} \!\! ,
\end{eqnarray}
where $k_z = \sqrt{k^2 - k_\parallel^2}$ for $k_\parallel < k$, $k_z = i\sqrt{k_\parallel^2 - k^2}$ for $k_\parallel > k$ and $r^p$ and $r^s$ are the Fresnel reflection coefficients for p-polarized and s-polarized waves, respectively. For a perfect reflector, the Fresnel coefficients are given by $r^p = 1$ and $r^s = -1$. Taking the imaginary part of the previous equation and performing the integration in $k_\parallel$, we obtain
\begin{eqnarray}
\mathrm{Im}\mathbb{G}_{xx}(\omega) &=& \mathrm{Im}\mathbb{G}_{yy}(\omega) = \frac{\omega}{4\pi c} \times \cr\cr
&\times &\!\!\!\!\!\left[ - \frac{\sin(2kz)}{(2kz)} - \frac{\cos(2kz)}{(2kz)^2} + \frac{\sin(2kz)}{(2kz)^3}\right] \!\! ,
\\
\mathrm{Im}\mathbb{G}_{zz}(\omega) &=& \frac{\omega}{2\pi c}\left[ - \frac{\cos(2kz)}{(2kz)^2} + \frac{\sin(2kz)}{(2kz)^3}\right].
\end{eqnarray}
Plugging previous expressions into Eq. \eqref{GammaGreen} and identifying the integrand as the spectral density function, we recover the result given by equation \eqref{gamma-1placa}.

Though the two methods are equivalent, it is worth noting that to identify the angular distribution of the emitted photons by using the Green's function method is not  an easy task as it is in the framework of  the field modes approach. Besides, although the calculation of the TPSE rate by using the Green's function method is in principle straightforward, this procedure may obscure the basic underlying physics of the problem. For some systems, the field modes approach may even emphasize  some physical aspects that are not evident in the Green's function method, as for instance when the system supports different types of modes and the emitter might de-excite by different pathways.

\section{Relation between the TPSE and the Purcell factors}

Now, we shall relate the TPSE spectral density function of an emitter near a surface to the one-photon SE rate in the same situation. In a previous work, this has been shown by considering the emitter near a planar interface\cite{rivera2017}. In the following derivation we shall not assume this restriction and shall consider the emitter near a surface with an arbitrary shape. This is possible by noting that $\mathrm{Im}\mathbb{G}(\omega)$ is a real and symmetric matrix\cite{buhmann2012}, which means it can be diagonalized. For systems where the basis which diagonalizes the imaginary part of the Green's function does not depend on frequency, one can write
\begin{equation}
\frac{\gamma(\omega;{\bf r})}{\gamma_0(\omega)} = 
\sum_{I,J}\frac{|\mathbb{D}_{IJ}(\omega,\omega_{eg} - \omega)|^2}{|\mathbb{D}(\omega,\omega_{eg} - \omega)|^2}
P_I(\omega;{\bf r})P_J(\omega_{eg} - \omega;{\bf r}), \label{gamma/gamma_0}
\end{equation}
where we defined the Purcell factors $P_I$'s, $I=1,2,3$, as
\begin{equation}
P_I(\omega;{\bf r}) := \frac{6\pi c}{\omega} \mathrm{Im}\mathbb{G}_{II}(\omega; {\bf r}, {\bf r}). \label{PurcellFactors}
\end{equation}
Note that the Purcell factors coincide with the ratio \eqref{Gamma1Green} if we choose ${\bf \hat{n}}_{eg} = {\bf \hat{e}}_I({\bf r})$, i.e., the transition dipole moment oriented along one of the basis vectors. It is important to note that when the basis which diagonalizes the Green's function is frequency-dependent Eq. \eqref{gamma/gamma_0} becomes inappropriate. On the other hand, this equation establishes an explicit relation between TPSE and one-photon SE, hence, showing in a very clear way the dependence of the TPSE rate on the local density of states (LDOS). 
\subsection{An emitter near a half-space dielectric medium}
Using the Purcell factors relation just presented, we shall determine as an example the spectral enhancement of the TPSE of an emitter near a semi-infinite homogeneous dielectric dispersive medium ($z < 0$). It is clear that Eq. \eqref{gamma/gamma_0} gives a straightforward way to compute $\gamma(\omega;{\bf r})$ since, in this situation, the cartesian basis diagonalizes the Green's function and we already know the corresponding formulas for the one-photon SE rates. Moreover, as before, we shall consider only $s \rightarrow s$ transitions. The corresponding spectral enhancement is given by equation (\ref{gamma/gamma_0-transicao-s}) with the following Purcell factors\cite{novotny2012}
\begin{eqnarray}
P_1 = P_2 \!\! &=& \!\! 1 + \frac{3}{4} \int_0^{k} d \kappa_\parallel \, \frac{\kappa_\parallel}{k^3\xi} \mathrm{Re}[(k^2r^s(\kappa_\parallel) - \xi^2r^p(\kappa_\parallel))e^{2i\xi z}] \cr
&+& \frac{3}{4} \!\! \int_{k}^\infty \!\!\!\!\! d \kappa_\parallel \, \frac{\kappa_\parallel}{k^3\zeta} \mathrm{Im}[k^2r^s(\kappa_\parallel) + \zeta^2r^p(\kappa_\parallel)]e^{-2\zeta z} ,
\\
P_3 \!\! &=& \!\! 1 + \frac{3}{2} \int_0^{k}\!\! d\kappa_\parallel \, \frac{\kappa_\parallel^3}{k^3\xi} \mathrm{Re}\left[r^p(\kappa_\parallel)e^{2i\xi z}\right] \cr
&+& \frac{3}{2} \int_{k}^\infty\!\!\!\! d \kappa_\parallel \, \frac{\kappa_\parallel^3}{k^3\zeta} \mathrm{Im}\left[r^p(\kappa_\parallel)\right]e^{-2\zeta z},
\end{eqnarray}
where $\xi = \sqrt{k^2 - \kappa_\parallel^2}$, $\zeta = \sqrt{\kappa_\parallel^2 - k^2}$ and $r^s$ and $r^p$ are the Fresnel coefficients for s-polarized  and 
p-polarized waves, respectively. We have two important regimes, the far-field regime ($kz \gtrsim 1$) and the near-field regime ($kz << 1$). In the latter, the Purcell factors are very high in comparison to the former due to the coupling with evanescent modes. Here, we restrict ourselves to the near-field regime. Evoking the Lorentz model for dielectrics\cite{zangwill2013}, we expect a strong dependence of the LDOS on the dielectric resonance frequencies, as a dielectric reflects like a metal for waves with frequencies close to the resonances. Hence, applying Eq. \eqref{gamma/gamma_0} we expect a substantial enhancement in the spectral density function near the resonance frequencies.

\begin{figure}[h!]
\begin{center}
\includegraphics[width=0.8\linewidth]{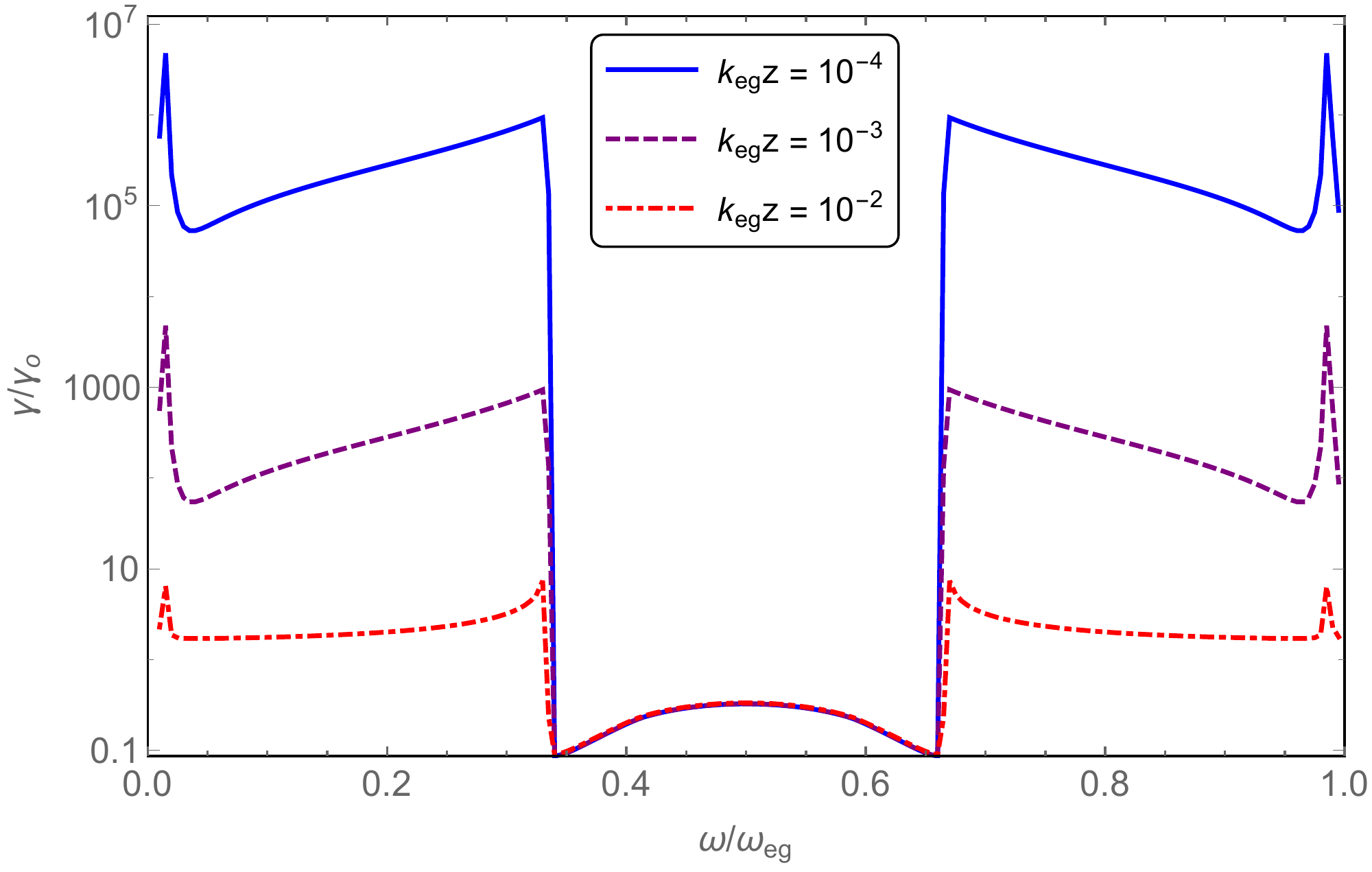}
\caption{Spectral enhancement of an emitter near a half-space Polystyrene medium as a function of $\omega/\omega_{eg}$ for three given values of $z$.  The Polystyrene resonance frequencies are given by ${\omega_{R1} = 5.54\times 10^{14}}$ rad/s and $\omega_{R2} = 1.35\times 10^{16}$ rad/s and the corresponding widths by $\Gamma = 1 \times 10^{11}$ rad/s\cite{hough1980}. The emitter transition frequency was chosen as $3\omega_{R2}$.}
\label{dieletrico}
\end{center}
\end{figure}

In Figure  \ref{dieletrico} we plot the spectral enhancement for three given distances between the emitter and the dielectric for an $s \rightarrow s$ transition. We considered a Polystyrene medium described by a Lorentz model with two resonance frequencies. The adjusted parameters were taken from reference\cite{hough1980}. First, note the huge changes on the spectral density function near the resonances frequencies $\omega_{R1}$ and $\omega_{R2}$. Since $\Gamma << \omega_{R1}, \omega_{R2}$, the transition between the two frequency regimes (before and after each resonance) occurs in a very narrow interval. Due to the symmetry of the spectral distribution with respect to $\omega_{eg}/2$, the same behavior occurs near the complementary frequencies $\omega_{eg} - \omega_{R1}$ and $\omega_{eg} - \omega_{R2}$. As it is evident from the figure, the spectral distribution may be orders of magnitude greater than its free space value, since we are in the near field regime. Furthermore, as the frequency of the emitted photon approaches a given resonance frequency from below, the spectral density increases monotonically until it reaches a maximum value, and then suffers an abrupt decrease as it crosses the resonance frequency. This result may open a possibility of controlling the spectral distribution of TPSE by tuning the resonance frequencies of a medium.
 %In analogy with the one-photon SE, a high LDOS at a given frequency $\omega_0$ favors the TPSE  at frequencies $\omega_0$ and $\omega_{eg} - \omega_0$.
%
\section{TPSE of an emitter between two parallel mirrors}
The suppression of spontaneous emission is a remarkable phenomenon which opens the possibility of manipulating excited atoms for large time intervals.
However, what is known as suppression of spontaneous emission is, in fact, the suppression of the dipolar one-photon emission. As the TPSE is a second order process, the ratio between the lifetimes of an emitter which decays by two-photon emission and by the emission of a single photon is about $10^8$. Hence, 
when the one-photon SE is supressed, in practice, the atom has an infinite lifetime.

The suppression of the one-photon SE is achieved when the partial LDOS vanishes at the transition frequency. Without a one-photon decay channel, in principle, the emitter can decay by the emission of a photon pair (in this work we are ignoring magnetic dipole transitions, quadrupolar transitions and so on). The suppression of the one-photon SE was first predicted by Barton\cite{barton1970} and observed only 15 years later by Hulet \textit{et al}\cite{hulet1985}. They considered atoms going through two perfectly conducting plates and also prepared the atoms so that they had a transition electric dipole moment parallel to the plates. They showed that, for distances between the plates smaller than half the transition wavelength, the atom does not decay by the emission of a single photon. 

In this section we investigate the TPSE in the same situation, namely, with the atom between two conducting parallel plates, by using the Purcell factors relation presented in the previous section. First, we recall that the Purcell factors in the cartesian basis are given by\cite{milonni2013,barton1970}
\begin{eqnarray}
P_1 = P_2 \!\! &=& \!\! \frac{3\pi}{2k L} \sum_{n = 0}^{\left[kL/\pi\right]}\sin^2\left(\frac{n\pi z}{L}\right) \left[ 1  + \frac{n^2\pi^2}{k^2L^2}\right], \label{Pparallel-2placas}
\\
P_3 \!\! &=& \!\! \frac{3\pi}{kL}\sum_{n = 0}^{\left[kL/\pi\right]}\cos^2\left(\frac{n\pi z}{L}\right)\left[1 - \frac{n^2\pi^2}{k^2 L^2}\right], \label{Pperp-2placas}
\end{eqnarray}
where $z$ is the distance between the atom and the first mirror (located at $z=0$), $L$ is the distance between the two mirrors and $\left[kL/\pi\right]$ means the greatest integer smaller than $kL/\pi$. We notice that $P_1$ and $P_2$ vanish for $L < \pi/k$, which means that the one-photon emission will be suppressed if the transition dipole moment is parallel to the plates. However, Eq. \eqref{gamma/gamma_0} shows that $\gamma$ also depends on $P_3$, which increases for small values of $L$. Therefore, it is clear that the TPSE is not completely suppressed in this situation unless $\mathbb{D}_{33} = 0$. For $s \rightarrow s$ and $d \rightarrow s$ transitions, which are the most common two-photon transitions, this can not be true due to the general form of $\mathbb{D}$\cite{florescu1987}. Consequently, for these type of transitions complete suppression of the TPSE can never occur for an emitter between two parallel conducting plates.

\begin{figure}[h!]
\begin{center}
\includegraphics[width=0.8\linewidth]{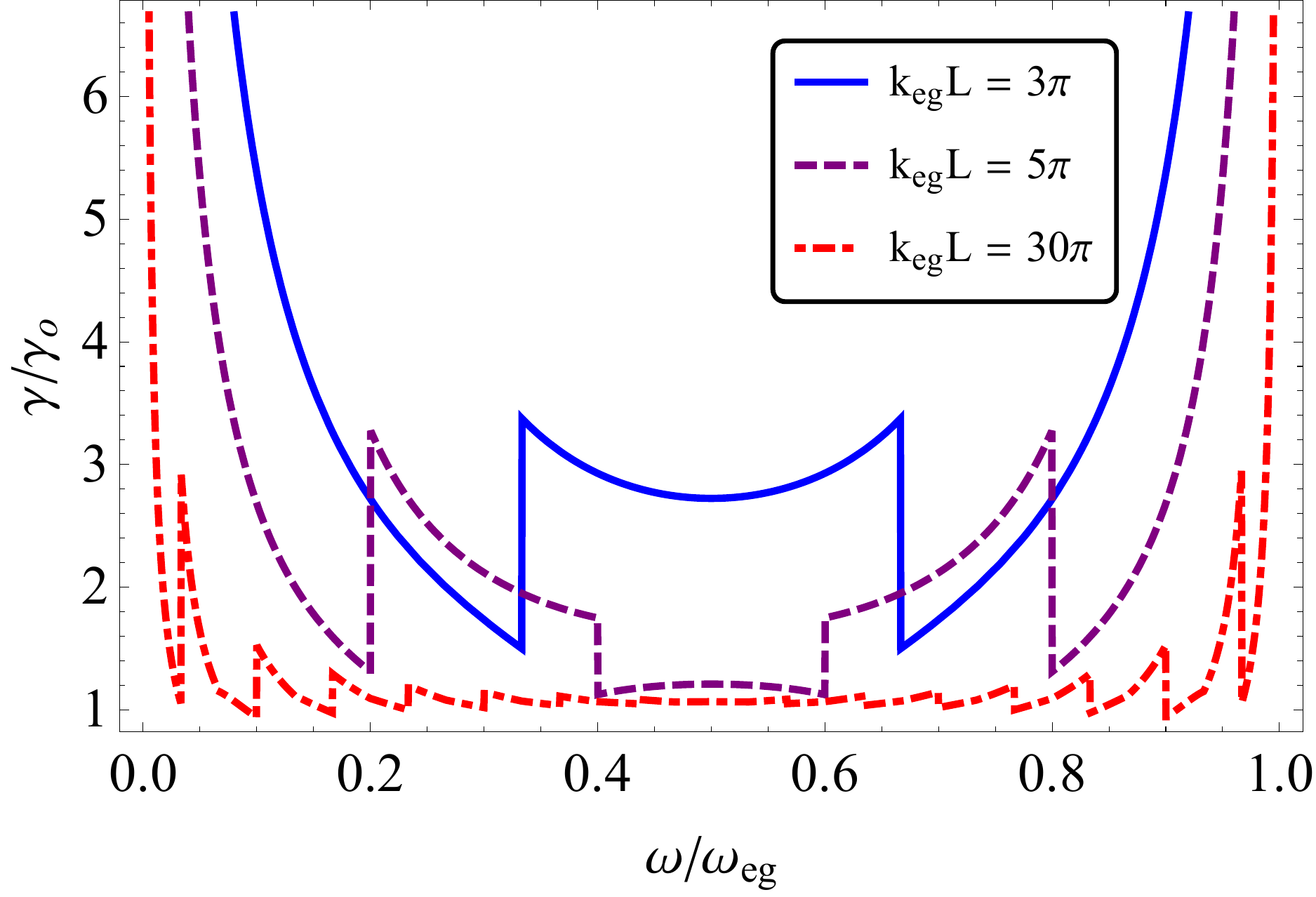}
\caption{Spectral enhancement of an emitter equidistant from two perfect mirrors as a function of $\omega/\omega_{eg}$ for three given values of $L$. }
\label{2placafrequencia}
\end{center}
\end{figure}

In Figure \ref{2placafrequencia} we plot the spectral enhancement for a ${s \rightarrow s}$ transition with the emitter equidistant from both plates, $\gamma(\omega,z=L/2)/\gamma_0(\omega)$, as a function of $\omega/\omega_{eg}$ for different values of $L$. Observe that complete suppression never  occurs. Note also that the spectral enhancement diverges for $\omega \rightarrow 0$ and $\omega \rightarrow \omega_{eg}$. This is a consequence of the fact that both $\gamma(\omega; z = L/2)$ and $\gamma_0(\omega)$ go to zero at these frequencies, but with different power laws. Further, note that as $L$ increases the spectral density function tends to the free space spectral density function (dotted-dashed line), as expected, since in this case the plates do not influence the emitter anymore.

\begin{figure}[h!]
\begin{center}
\includegraphics[width=0.9\linewidth]{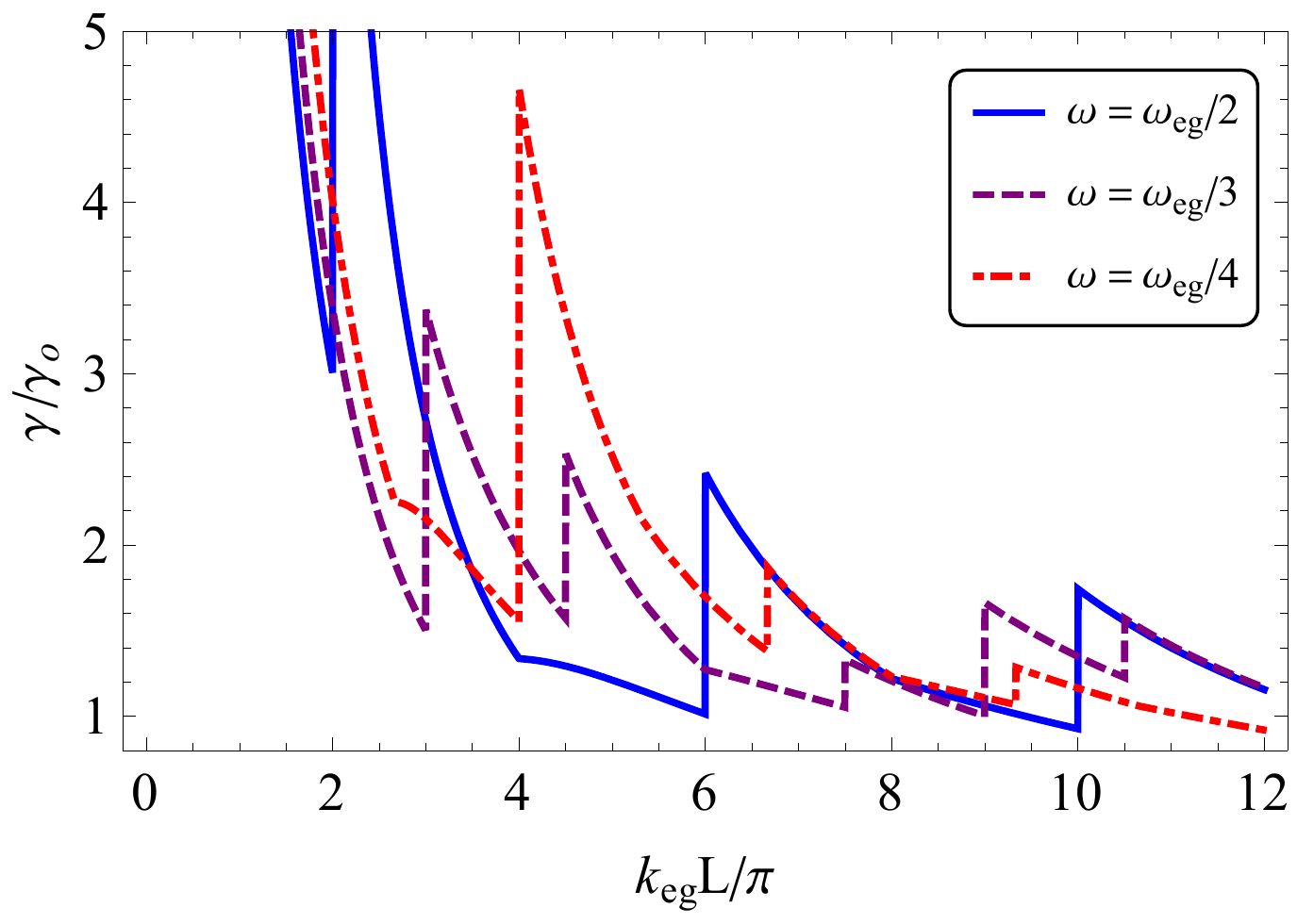}
\caption{Spectral enhancement in terms of the dimensionless variable $k_{eg}L/\pi$ with the  emitter equidistant to both mirrors for three given frequencies.}
\label{2placadistancia}
\end{center}
\end{figure}

In Figure \ref{2placadistancia} we plot the spectral enhancement with the emitter equidistant from both plates, $\gamma(\omega,z=L/2)/\gamma_0(\omega)$, as a function of $L$ for three given  values of frequency. In analogy to what happens in the one-photon SE, as the distance between the plates crosses certain multiples of $kL/\pi$, the number of available modes changes abruptly, giving rise to the discontinuities in the spectral density function. However, contrary to what happens with the one-photon SE rate when the transition dipole moment is parallel to the plates, total suppression never occurs.

The non-suppression of the TPSE in this situation is a consequence of the fact that the spectral density function is not proportional to the partial LDOS, in contrast to what happens in the one-photon SE\cite{novotny2012}. As we can see from Eq. \eqref{diadico}, the TPSE rate does not depend explictly on the transition dipole moment, but on the intermediate transition dipole moments. 
However, the enhancement or suppression of the TPSE at a given frequency, say $\omega_1$, can occur. It suffices that the LDOS is enhanced or vanishes at this frequency (or at  the complementary frequency $\omega_{eg}- \omega_0$). 
%The same can be said for the enhancement of the TPSE at a given frequency.
%
\section{Conclusions and final remarks}
In this work we presented an alternative formula for computing the TPSE rate of an excited emitter near a surface of arbitrary shape which is written in terms of the electromagnetic field modes. As a check of self-consistency, we used this formula to reobtain the TPSE rate in free space as well as with the emitter near a perfectly conducting plate. We showed explicitly the equivalence of our formula with the one usually found in the literature  which is written in terms of the dyadic Green's function, Eq.  \eqref{GammaGreen}. We did that in a simple and straightforward way, providing an alternative demonstration of Eq. \eqref{GammaGreen}. We compared both methods and showed that, although the TPSE rate calculation by using the Green's function method is straightforward, the field modes approach has the advantage of clarifying the physics of the problem and provides an easier way to compute the angular distribution of  the emitted photons.
%
%Using this formula, we performed a demonstration of equation \eqref{GammaGreen} by a very simple method, which is different from what can be found in literature.
%
 We have also related the TPSE spectral density function of an emitter near an arbitrary object to the corresponding one-photon SE Purcell factors. This allowed us to identify the general dependence of the TPSE rate on the field LDOS and also provided us a very simple way to compute the spectral density function in any situation where the one-photon SE rate is known. We applied this formalism for an emitter near a homogeneous semi-infinite dielectric medium and verified that an interesting behaviour of the TPSE spectral distribution shows up near the resonance frequencies of the dielectric. This result opens the possibility of controlling the frequencies of the two emitted photons by tuning the resonance frequencies of the medium through an external agent.
 %
 %peaks in the LDOS at a given frequency increases the TPSE at the same frequency (and at the complementary frequency). 
 %
 Finally, we analyzed the TPSE of an emitter between two parallel  perfect mirrors and showed that the TPSE can not be completely suppressed for $s \rightarrow s$ transitions. This is to be contrasted to the suppression of the one-photon SE that may occur for an atom between two parallel perfect mirrors if the atom is appropriately prepared\cite{hulet1985}. 
 %We hope that this advance in the conceptual understanding of TPSE enlighten future works in this field.
%
\section{Acknowledgements}
 The authors thank to L. Davidovich and M. Lima for enlightening discussions. Y.M., D.S., F.S.S.R. and C.F. acknowledge the brazilian agencies CAPES, CNPq and FAPERJ for partially financing this research. W.K.-K thanks LANL LDRD program for financial support.
\section{Appendix}
\subsection{Transition between isotropic states}
In this appendix we investigate the functional form of $\mathbb{D}$ when the initial and final states are $s$ states. This was already investigated in previous works\cite{breit1940,florescu1987} even for others transitions such as $d \rightarrow s$. Considering the one electron states $|e\rangle = |n_e,l = 0, m = 0\rangle$ and $|g\rangle = |n_g,l = 0, m = 0\rangle$, the intermediate states which the matrix elements of ${\bf d}$ do not vanish are $p$ states and can be written as $|k\rangle = |n,l = 1, m = 0, \pm 1\rangle$. The corresponding wave functions can be written as a product of a radial function by the spherical harmonics. We have $\psi_e({\bf r}) =  R_{n_e0}(r)Y_{00}$, $\psi_g({\bf r}) =  R_{n_g0}(r)Y_{00}$ and $\psi_{nm}({\bf r}) =  R_{n1}(r)Y_{1m}(\theta,\phi)$, so the intermediate transition dipole moments are given by
\begin{eqnarray}
{\bf d}_{e,nm} =  eY_{00}^\ast\int_0^\infty \!\!\!\!\!\! dr r^3R_{n_e0}^\ast(r)R_{n1}(r)\int \!\! d\Omega {\bf \hat{r}}Y_{1m}(\theta,\phi),
\\
{\bf d}_{g,nm} =  eY_{00}^\ast\int_0^\infty \!\!\!\!\!\! dr r^3R_{n_g0}^\ast(r)R_{n1}(r)\int \!\! d\Omega {\bf \hat{r}}Y_{1m}(\theta,\phi).
\end{eqnarray}
From these expressions, we note that the directions of these vectors depend only on the angular integrals, while their modula depend on the quantum numbers $n_e$, $n_g$ and $n$. As $l$ is fixed, the directions depend only on $m$. Using the expansion
\begin{equation}
{\bf \hat{r}} = \sqrt{\frac{4\pi}{3}}\left\lbrace \frac{\left( Y_{1-1} - Y_{11}\right)}{\sqrt{2}}{\bf \hat{x}} + i\frac{\left( Y_{1-1} + Y_{11}\right)}{\sqrt{2}}{\bf \hat{y}} + Y_{10}{\bf \hat{z}}\right\rbrace ,
\end{equation}
and calculating explicitly the angular integral for all values of $m$, we obtain
\begin{eqnarray}
{\bf d}_{e,nm} =  d_{en}\hat{\boldsymbol{\epsilon}}_{m},
\\
{\bf d}_{g,nm} =  d_{gn}\hat{\boldsymbol{\epsilon}}_{m},
\end{eqnarray}
where the set $\left\lbrace\hat{\boldsymbol{\epsilon}}_{-1},\hat{\boldsymbol{\epsilon}}_0,\hat{\boldsymbol{\epsilon}}_{1}\right\rbrace$ is an orthonormal basis in three dimensions given by
\begin{equation}
\hat{\boldsymbol{\epsilon}}_{-1} = -\frac{({\bf \hat{x}} + i{\bf \hat{y}})}{\sqrt{2}}\, ;\;\;\; 
\hat{\boldsymbol{\epsilon}}_{1} = \frac{{\bf \hat{x}} - i{\bf \hat{y}}}{\sqrt{2}}\, ;\;\;\;
\hat{\boldsymbol{\epsilon}}_{0} = {\bf \hat{z}}.
\end{equation}
With this result, we obtain
\begin{equation}
\mathbb{D}(\omega_{\alpha},\omega_{\alpha'}) = \! \sum_nd_{en}d_{ng}\! \left[\frac{1}{\omega_{en} - \omega_{\alpha}} + \frac{1}{\omega_{en} - \omega_{\alpha}}\right]\!\! \sum_m \hat{\boldsymbol{\epsilon}}_{m}\hat{\boldsymbol{\epsilon}}_{m}^\ast .
\end{equation}
Due to the fact that the vectors $\hat{\boldsymbol{\epsilon}}_{m}$ form a basis, we have $\sum_m \hat{\boldsymbol{\epsilon}}_{m}\hat{\boldsymbol{\epsilon}}_{m}^\ast = \mathbb{I}$ and $\mathbb{D}$ takes the form $\mathbb{D}(\omega_{\alpha},\omega_{\alpha'}) = D(\omega_{\alpha},\omega_{\alpha'})\mathbb{I}$, where
\begin{equation}
D(\omega_{\alpha},\omega_{\alpha'}) = \sum_nd_{en}d_{ng}\left[\frac{1}{\omega_{en} - \omega_{\alpha}} + \frac{1}{\omega_{en} - \omega_{\alpha'}}\right].
\end{equation}
Finally, we can rewrite equations \eqref{GammaModos}, \eqref{GammaGreen} and \eqref{gamma/gamma_0} respectively as
\begin{eqnarray}
\Gamma({\bf r}) &=& \frac{\pi}{4\epsilon_0^2\hbar^2}\sum_{\alpha,\alpha'}\omega_{\alpha}\omega_{\alpha'}|D(\omega_{\alpha},\omega_{\alpha'})|^2 \times \cr
&\times & |{\bf A}_{\alpha}({\bf r})\cdot {\bf A}_{\alpha'}({\bf r})|^2\delta(\omega_{\alpha} + \omega_{\alpha'} -\omega_{eg}) , \label{GammaModos-transicao-s}
\\
\Gamma({\bf r}) &=& \frac{\mu_0^2}{\pi\hbar^2}\int_0^{\omega_{eg}}d\omega \omega^2(\omega_{eg} - \omega)^2|D(\omega,\omega_{eg} - \omega)|^2 \times \cr
&\times & \mathrm{Tr}[\mathrm{Im}\mathbb{G}(\omega;{\bf r},{\bf r})\cdot \mathrm{Im}\mathbb{G}(\omega_{eg} - \omega;{\bf r},{\bf r})] , \label{GammaGreen-transicao-s}
\\
\frac{\gamma(\omega,{\bf r})}{\gamma_0(\omega)} &=& \frac{1}{3}\sum_{I}P_I(\omega,{\bf r})P_I(\omega_{eg} - \omega,{\bf r}).
\end{eqnarray}
\newpage
%%%%%%%%%%%%%%%%%%%%%%%%%%% Bibliografia %%%%%%%%%%%%%%%%%%%%%%%%%%%%%%%%

\bibliography{references} 
\bibliographystyle{ieeetr}

%%%%%%%%%%%%%%%%%%%%%%%%%%%%%%%%%%%%%%%%%%%%%%%%%%%%%%%%%%%%%%%%%%%%%%%%%
%

\end{document}